\documentstyle[aps]{revtex}
\tighten

\begin{document}

\draft

%======================================%
%<<<<<<<<<<<< TITLE PAGE >>>>>>>>>>>>>>%
%======================================%

\preprint{hep-th/0006146}
\title{Perturbation of junction condition and doubly gauge-invariant
variables} 
\author{Shinji Mukohyama}
\address{
Department of Physics and Astronomy, University of Victoria\\ 
Victoria, BC, V8W 3P6, Canada
}
\date{\today}

\maketitle

%======================================%
%<<<<<<<<<<<<< ABSTRACT >>>>>>>>>>>>>>>% 
%======================================%

\begin{abstract} 

The junction condition across a singular surface in general
relativity, formulated by Israel, has double covariance. 
In this paper, a general perturbation scheme of the junction condition
around an arbitrary background is given in a doubly covariant
way. After that, as an application of the general scheme, we consider
perturbation of the junction condition around a background with the
symmetry of a $(D-2)$-dimensional constant curvature space, where $D$
is the dimensionality of the spacetime. The perturbed junction
condition is written in terms of doubly gauge-invariant variables
only. Since the symmetric background includes cosmological solutions
in the brane-world as a special case, the doubly gauge-invariant
junction condition can be used as basic equations for perturbations in
the brane-world cosmology. 

\end{abstract}


%======================================%
%<<<<<<<<<< Introduction  >>>>>>>>>>>>>%
%======================================%

\section{Introduction}

%%%%% Brane-world

Brane-world scenario proposed by Randall and Sundrum~\cite{RS} has
been attracting a lot of interests. In this scenario, our
$4$-dimensional universe is considered as a world volume of a
$3$-brane, or a timelike hypersurface, in $5$-dimensional bulk
spacetime. It was shown that, in a $5$-dimensional anti de-Sitter
background, gravity on the $3$-brane can be effectively described by
$4$-dimensional Newton's law if the tension of the $3$-brane is
fine-tuned. Since this scenario may be realistic and may give drastic
changes to our understanding of $4$-dimensional gravity, many works
have been done from various points of view. For example,
$4$-dimensional effective Einstein equation on the $3$-brane was
derived~\cite{SMS}; gravitational perturbations in the Randall-Sundrum
background were discussed~\cite{CG,GT,Tanaka-Montes,SSM,GKR,Tanaka};
black holes in the brane-world were discussed~\cite{CHR,EHM};
cosmology based on this scenario was also
discussed~\cite{FTW,BDEL,Mukohyama1,Kraus,Ida,MSM,GS,Koyama-Soda,HHR}. 
Recently, formalisms to treat perturbations in the brane-world
cosmology were
proposed~\cite{Mukohyama2,KIS,Maartens,Langlois,BDBL,KS,LMW}. In
particular, gauge-invariant master equations for perturbations in bulk
was obtained in ref.~\cite{Mukohyama2}, and the corresponding
perturbed junction condition was obtained in ref.~\cite{KIS}.

%%%%% Junction condition

In most of works on the brane-world scenario, the covariant junction
condition formulated by Israel~\cite{Israel} is used to treat a
singular surface, or the world volume of the $3$-brane. The junction
condition relates a jump of extrinsic curvature of the singular
surface to a surface energy momentum tensor~\footnote{
An equivalent condition can be derived from an action principle with a
delta function source. See, for example, \cite{Hayward}.}. 
One of its advantages is the manifest double covariance of the 
formalism. Coordinates outside the singular surface and those
intrinsic to the singular surface can be completely independent, and
the formalism is covariant under coordinate transformations both 
outside and on the singular surface. The double covariance of the
junction condition is actually convenient for the purpose of the
analysis of the brane-world scenario: because of the double
covariance, coordinates intrinsic to our universe can be disentangled
from those in the higher dimensional spacetime and covariance in our
universe will be manifestly realized.

However, as far as the author knows, there is no manifestly doubly
covariant scheme of perturbation of the junction condition. 
Although perturbations of the junction condition around symmetric
backgrounds were analyzed by some
authors~\cite{Gerlach-Sengupta,KIF,II,KIS,Langlois,BDBL,KS}, their
formalism is not manifestly doubly covariant: coordinates intrinsic to
the singular surface are entangled with those outside.

%%%%% Purpose of this paper

The main purpose of this paper is to formulate a general perturbation
scheme of the junction condition around an arbitrary background in a
doubly covariant way. After that, as an application of the general
scheme, we consider perturbation of the junction condition around a
background with the symmetry of a $(D-2)$-dimensional constant
curvature space, where $D$ is the dimensionality of the spacetime. The
perturbed junction condition is written in terms of doubly
gauge-invariant variables only. The final expression of the doubly
gauge-invariant junction condition is equivalent to that in
ref.~\cite{KIS}. Without the doubly covariant formalism developed in
this paper, we could not judge whether the expression is actually
doubly gauge-invariant or not. Since the symmetric background includes 
cosmological solutions in the brane-world as a special case, the
doubly gauge-invariant junction condition can be used as basic
equations for perturbations in the brane-world cosmology. The
gauge-invariance with respect to coordinates intrinsic to the singular 
surface, or the brane, is convenient since observables for observers
on the brane should have such gauge-invariance. The gauge-invariance
with respect to coordinates in the higher dimensional spacetime, or
bulk, must be made sure since it is expected that a change of these
coordinates does not affect physics on the brane. Therefore, the
doubly gauge-invariant junction condition is expected to be useful in
the analysis of perturbations in the brane-world cosmology.

%%%%% Organization of this paper

In Sec.~\ref{sec:junciton-cond} the junction condition formulated by 
Israel is reviewed in a doubly covariant form. 
In Sec.~\ref{sec:general} a general scheme of perturbation of the
junction condition is given. 
In Sec.~\ref{sec:symmetric-back}, as an application of the general
scheme, we consider a symmetric background and the doubly gauge-invariant
junction condition is derived. 
Sec.~\ref{sec:summary} is devoted to a summary of this paper and
discussions.

%======================================%
%<<<<<<<< Junction condition >>>>>>>>>>%
%======================================%

\section{Junction condition in a doubly covariant form}
	\label{sec:junciton-cond}

%%%%% Inbedding of a hypersurface

Let us consider a $D$-dimensional spacetime ${\cal M}$ and a spacelike
or timelike hypersurface $\Sigma$ which separates ${\cal M}$ into two
regions ${\cal M}_+$ and ${\cal M}_-$. Suppose that in each region the
inbedding of the hypersurface $\Sigma$ is specified by the following
parametric equations 
%============< EQUATION >==============%
%
\begin{equation}
 \Sigma: x^M_{\pm} = Z_{\pm}^M(y), 
\end{equation}
%======================================%
where $\{x^M_{\pm}\}$ denote spacetime coordinates in the region
$M_{\pm}$, respectively, and $y$ denotes the set $\{y^{\mu}\}$ of
$D-1$ parameters corresponding to intrinsic coordinates of 
$\Sigma$. In order to make $\Sigma$ well-defined the two sets of 
functions $\{Z_{+}^M(y)\}$ and $\{Z_{-}^M(y)\}$ should transform as 
%============< EQUATION >==============%
%
\begin{equation}
 Z_{\pm}^M(y) \to 
	{Z'}_{\pm}^M(y)=F_{\pm}^M(Z_{\pm}(y))
	\label{eqn:tr-Z}
\end{equation}
%======================================%
under the $D$-dimensional coordinate transformations 
%============< EQUATION >==============%
%
\begin{equation}
 x_{\pm}^M \to {x'}_{\pm}^M = F_{\pm}^M(x_{\pm}), 
	\label{eqn:coord-tr}
\end{equation}
%======================================%
respectively. We shall call these coordinate transformations {\it
$D$-coordinate transformations}. The $D$-coordinate transformations
can be taken independently in the regions ${\cal M}_+$ and 
${\cal M}_-$. Hereafter, in most cases, we shall omit the sign $\pm$
which distinguishes quantities in ${\cal M}_+$ and those in 
${\cal M}_-$.

%%%%% Induced metric and extrinsic curvature

The induced metric on $\Sigma$ is 
%============< EQUATION >==============%
%
\begin{equation}
 q_{\mu\nu}\equiv e^M_{\mu}e^N_{\nu}g_{MN}, 
\end{equation}
%======================================%
where $g_{MN}$ is the spacetime metric and $\{e^M_{\mu}\}$ are
tangential vectors defined by 
$e^M_{\mu}\equiv\partial Z^M/\partial y^{\mu}$. 
Since $\partial_{[\mu}\partial_{\nu]}Z^M(y)=0$, we have the equation
$[e_{\mu},e_{\nu}]^M=0$. The extrinsic curvature of $\Sigma$ is
defined by 
%============< EQUATION >==============%
%
\begin{equation}
 K_{\mu\nu} = \frac{1}{2}e^M_{\mu}e^N_{\nu}{\cal L}_{n}g_{MN}, 
\end{equation}
%======================================%
where ${\cal L}$ represents the $D$-dimensional Lie derivative and $n$
is the unit normal of $\Sigma$ determined by 
%============< EQUATION >==============%
%
\begin{eqnarray}
 n_M e^M_{\mu} & = & 0, \nonumber\\
 n_Mn^M & = & \epsilon,
\end{eqnarray}
%======================================%
where $\epsilon=1$ for timelike $\Sigma$ and $\epsilon=-1$ for
spacelike $\Sigma$. 
Here, we mention that $e^M_{\mu}$ transforms 
as a $D$-vector under the $D$-coordinate transformation
(\ref{eqn:coord-tr}) because of the transformation
(\ref{eqn:tr-Z}). Hence, the induced metric $q_{\mu\nu}$ and the 
extrinsic curvature $K_{\mu\nu}$ transform as $D$-scalars under the
$D$-coordinate transformation (\ref{eqn:coord-tr}). Therefore, they
are invariant under the $D$-coordinate transformation, provided that
they are considered as functions of $y$. On the other hand, they
transform as $(D-1)$-tensors under the reparameterization of $\Sigma$ 
%============< EQUATION >==============%
%
\begin{equation}
 y^{\mu} \to {y'}^{\mu} = f^{\mu}(y). 
	\label{eqn:reparametrization}
\end{equation}
%======================================%
We shall call this reparameterization {\it $(D-1)$-coordinate
transformation} since $\{y^{\mu}\}$ can be considered as coordinates
in the $(D-1)$-dimensional manifold $\Sigma$. It seems worth while
stressing that the $(D-1)$-coordinate transformation is NOT a part of
the $D$-coordinate transformation. They are completely independent.

%%%%% Junction condition

Since the intrinsic geometry of $\Sigma$ should be regular, the
induced metric calculated from one side and another should be 
identical:
%============< EQUATION >==============%
%
\begin{equation}
 q_{\mu\nu +} = q_{\mu\nu -} \equiv q_{\mu\nu}. 
\end{equation}
%======================================%
Then, the junction condition formulated by Israel~\cite{Israel}
relates the jump of the extrinsic curvature to the surface energy
momentum tensor $S_{\mu\nu}$ associated with $\Sigma$ as 
%============< EQUATION >==============%
%
\begin{equation}
 K_{\mu\nu +} - K_{\mu\nu -} = 
	-\kappa^2 \left(S_{\mu\nu}-\frac{1}{D-2}S q_{\mu\nu}\right), 
\end{equation}
%======================================%
where $S=q^{\mu\nu}S_{\mu\nu}$, $q^{\mu\nu}$ is the inverse of the
induced metric $q_{\mu\nu}$, and $\kappa^2$ is the $D$-dimensional
gravitational coupling constant. Here, we have assumed that the unit
normal $n_M$ is directed from ${\cal M}_-$ to ${\cal M}_+$. Note that
the surface energy momentum tensor $S_{\mu\nu}$ is invariant under the
$D$-coordinate transformation, while it transforms as a $(D-1)$-tensor 
under the $(D-1)$-coordinate transformation. Since the junction
condition was first introduced by Lanczos~\cite{Lanczos} in a
non-covariant form, the surface energy momentum tensor $S_{\mu\nu}$ is 
often called Lanczos tensor.

%======================================%
%<<<<<<< General perturbation >>>>>>>>>%
%======================================%

\section{General perturbation}
	\label{sec:general}

%%%%% Perturbation of metric and inbedding relation

Now let us consider general perturbations of $g_{MN}$ and $Z^M(y)$
around an arbitrary background specified by $g_{MN}^{(0)}$ and
$Z^{(0)M}(y)$: 
%============< EQUATION >==============%
%
\begin{eqnarray}
 g_{MN} & = & g_{MN}^{(0)} + \delta g_{MN}, \nonumber\\
 Z^M(y) & = & Z^{(0)M}(y) + \delta Z^M(y). 
\end{eqnarray}
%======================================%
The unperturbed functions $Z^{(0)M}(y)$ determine the unperturbed
hypersurface $\Sigma^{(0)}$, while the perturbed functions $Z^M(y)$
determine the perturbed hypersurface $\Sigma$. Therefore, if we define 
$e^{(0)M}_{\mu}$ by $\partial Z^{(0)M}/\partial y^{\mu}$ then it is 
well-defined only on $\Sigma^{(0)}$, while $e^M_{\mu}$ is well-defined
only on $\Sigma$. Hence, in order to develop perturbation formalism,
it seems convenient to extend those unperturbed tangent vectors
$e^{(0)M}_{\mu}$ off $\Sigma^{(0)}$ so that both $e^M_{\mu}$ and
$e^{(0)M}_{\mu}$ become well-defined on $\Sigma$. The final
expressions of the perturbed junction condition will be independent of
the way of the extension. For the purpose of extension of
$e^{(0)M}_{\mu}$, we consider one-parameter family of hypersurfaces 
$\Sigma^{(0)}_{\varphi}$. 
%============< EQUATION >==============%
%
\begin{equation}
 \Sigma^{(0)}_{\varphi}:\ x^M = Z^{(0)M}_{\varphi}(y).
	\label{eqn:Sigma-varphi-def}
\end{equation}
%======================================%
We assume that $Z^{(0)M}_0(y)=Z^{(0)M}(y)$ so that
$\Sigma^{(0)}_0=\Sigma^{(0)}$. By using the one-parameter family of 
hypersurfaces, we can define a set of tangent vectors by 
%============< EQUATION >==============%
%
\begin{equation}
 e^{(0)M}_{\mu} = \partial Z^{(0)M}_{\varphi}/\partial y^{\mu}. 
\end{equation}
%======================================%
Since $\partial_{[\mu}\partial_{\nu]}Z^{(0)M}_{\varphi}(y)=0$, we have
the equation $[e^{(0)}_{\mu},e^{(0)}_{\nu}]^M=0$. 
With this definition, the perturbed tangent vectors are written as
follows up to the linear order. 
%============< EQUATION >==============%
%
\begin{equation}
 e^M_{\mu} = e^{(0)M}_{\mu} - {\cal L}_{\delta Z}e^{(0)M}_{\mu}. 
\end{equation}
%======================================%

\subsection{Perturbations of induced metric and extrinsic curvature} 

%%%%% Induced metric

Correspondingly, the induced metric of $\Sigma$ can be written up to
the linear order as
%============< EQUATION >==============%
%
\begin{equation}
 q_{\mu\nu} = \bar{q}^{(0)}_{\mu\nu} 
	+ e^{(0)M}_{\mu}e^{(0)N}_{\nu}\delta g_{MN}
	- {\cal L}_{\delta Z}(e^{(0)M}_{\mu}e^{(0)N}_{\nu})
	g^{(0)}_{MN},
	\label{eqn:q-barq}
\end{equation}
%======================================%
where
%============< EQUATION >==============%
%
\begin{equation}
 \bar{q}^{(0)}_{\mu\nu} = 
	e^{(0)M}_{\mu}e^{(0)N}_{\nu}g^{(0)}_{MN}. 
\end{equation}
%======================================%
However, $\bar{q}^{(0)}_{\mu\nu}$ in eq.~(\ref{eqn:q-barq}) is not the
unperturbed induced metric 
$q^{(0)}_{\mu\nu}(y)=\bar{q}^{(0)}_{\mu\nu}|_{x=Z^{(0)}(y)}$
on $\Sigma^{(0)}$ since the former should be evaluated on $\Sigma$ at 
$x^M=Z^M(y)$. Rather, they are related as  
%============< EQUATION >==============%
%
\begin{equation}
 \left.\bar{q}^{(0)}_{\mu\nu}\right|_{x=Z(y)} = 
	q^{(0)}_{\mu\nu}(y) + \delta Z^M(y)
	\left.\partial_M\bar{q}^{(0)}_{\mu\nu}\right|_{x=Z^{(0)}(y)},
\end{equation}
%======================================%
where we have used the fact that $\bar{q}^{(0)}_{\mu\nu}$ is a
$D$-scalar and Taylor-expanded it. Therefore, eq.~(\ref{eqn:q-barq})
can be written as $q_{\mu\nu}=q^{(0)}_{\mu\nu}+\delta q_{\mu\nu}$,
where 
%============< EQUATION >==============%
%
\begin{equation}
 \delta q_{\mu\nu} = e^{(0)M}_{\mu}e^{(0)N}_{\nu}
	(\delta g_{MN} + {\cal L}_{\delta Z}g^{(0)}_{MN}). 
	\label{eqn:delta-q}
\end{equation}
%======================================%
Note that $\delta q_{\mu\nu}$ may be evaluated on $\Sigma^{(0)}$
since the difference from the value on $\Sigma$ is the second order.

%%%%% Extrinsic curvature

The unit normal of $\Sigma$ is written up to the linear order as 
$n_M=n^{(0)}_M+\delta n_M$, where $n^{(0)}_M$ is determined by 
%============< EQUATION >==============%
%
\begin{eqnarray}
 e^{(0)M}_{\mu} n^{(0)}_M & = & 0, \nonumber\\
 n^{(0)M}n^{(0)}_M & = & \epsilon,
\end{eqnarray}
%======================================%
and $\delta n_M$ is given by 
%============< EQUATION >==============%
%
\begin{eqnarray}
 e^{(0)M}_{\mu}\delta n_M & = & 
	n^{(0)}_M{\cal L}_{\delta Z}e^{(0)M}_{\mu},
	\nonumber\\
  n^{(0)M}\delta n_M & = &
	\frac{1}{2}n^{(0)M}n^{(0)N}\delta g_{MN},
	\label{eqn:delta-n}
\end{eqnarray}
%======================================%
or 
%============< EQUATION >==============%
%
\begin{equation}
 \delta n^M = \frac{\epsilon}{2}n^{(0)M}
	n^{(0)N}n^{(0)L}\delta g_{NL}
	+ e^{(0)M}_{\mu}q^{(0)\mu\nu}n^{(0)}_N
	{\cal L}_{\delta Z} e^{(0)N}_{\nu}. 
\end{equation}
%======================================%
Here, $n^{(0)M}=g^{(0)MN}n^{(0)}_N$, 
$\delta n^M=g^{(0)MN}\delta n_N$ and $q^{(0)\mu\nu}$ is the inverse of 
$q^{(0)}_{\mu\nu}$. 
The extrinsic curvature of $\Sigma$ is 
%============< EQUATION >==============%
%
\begin{equation}
 K_{\mu\nu} = \bar{K}^{(0)}_{\mu\nu} 
	+ \frac{1}{2}e^{(0)M}_{\mu}e^{(0)N}_{\nu}
	({\cal L}_{\delta n}g^{(0)}_{MN}
	- 2n^{(0)L}\delta\Gamma_{LMN})
	- \frac{1}{2}{\cal L}_{\delta Z}
	(e^{(0)M}_{\mu}e^{(0)N}_{\nu})
	{\cal L}_{n^{(0)}}g^{(0)}_{MN}
	\label{eqn:K-barK}
\end{equation}
%======================================%
up to the linear order, where
%============< EQUATION >==============%
%
\begin{eqnarray}
 \bar{K}^{(0)}_{\mu\nu} & = & 
	\frac{1}{2}e^{(0)M}_{\mu}e^{(0)N}_{\nu}
	{\cal L}_{n^{(0)}}g^{(0)}_{MN}, \nonumber\\
 \delta\Gamma_{LMN} & = & \frac{1}{2}
	(\delta g_{LM;N}+\delta g_{LN;M}-\delta g_{MN;L}),
\end{eqnarray}
%======================================%
and the semicolon denotes the $D$-dimensional covariant derivative
compatible with the unperturbed metric $g^{(0)}_{MN}$. The quantity 
$\bar{K}^{(0)}_{\mu\nu}$ in eq.~(\ref{eqn:K-barK}) should be
evaluated on $\Sigma$ at $x^M=Z^M(y)$ and is related to the
unperturbed extrinsic curvature
$K^{(0)}_{\mu\nu}(y)=\bar{K}^{(0)}_{\mu\nu}|_{x=Z^{(0)}(y)}$ of
$\Sigma^{(0)}$ as follows. 
%============< EQUATION >==============%
%
\begin{equation}
 \left.\bar{K}^{(0)}_{\mu\nu}\right|_{x=Z(y)} = 
	K^{(0)}_{\mu\nu}(y)
	+ \delta Z^M(y)
	\left.\partial_M\bar{K}^{(0)}_{\mu\nu}\right|_{x=Z^{(0)}(y)},
\end{equation}
%======================================%
where we have used the fact that $\bar{K}^{(0)}_{\mu\nu}$ is a
$D$-scalar and Taylor-expanded it. Therefore, eq.~(\ref{eqn:K-barK}) 
can be rewritten as $K_{\mu\nu}=K^{(0)}_{\mu\nu}+\delta K_{\mu\nu}$,
where 
%============< EQUATION >==============%
%
\begin{equation}
 \delta K_{\mu\nu} = \frac{1}{2}e^{(0)M}_{\mu}e^{(0)N}_{\nu}
	({\cal L}_{\delta n}g^{(0)}_{MN}
	+ {\cal L}_{\delta Z}{\cal L}_{n^{(0)}}g^{(0)}_{MN}
	- 2n^{(0)L}\delta\Gamma_{LMN}). 
	\label{eqn:delta-K-formal}
\end{equation}
%======================================%
Note that $\delta K_{\mu\nu}$ may be evaluated on $\Sigma^{(0)}$
since the difference from the value on $\Sigma$ is the second order.

%%%%% Another expression of \delta K_{\mu\nu}

From the expression (\ref{eqn:delta-K-formal}) we see that it is
not necessary to extend $\delta Z^M$ off $\Sigma^{(0)}$. Nonetheless,
for explicit calculation of $\delta K_{\mu\nu}$, it turns out to be 
convenient to rewrite the expression into another form by extending
$\delta Z^M$ off $\Sigma^{(0)}$ as a $D$-vector. Off course, the value
of $\delta K_{\mu\nu}$ does not depend on the way of extension off 
$\Sigma^{(0)}$. The following is an alternative expression of 
$\delta K_{\mu\nu}$. 
%============< EQUATION >==============%
%
\begin{eqnarray}
 \delta K_{\mu\nu} & = &
	\frac{\epsilon}{2}n^{(0)M}n^{(0)N}
	(\delta g_{MN}+2\delta Z_{M;N})	K^{(0)}_{\mu\nu}	
	\nonumber\\
 & & 	- \frac{1}{2}n^{(0)L}e^{(0)M}_{\mu}e^{(0)N}_{\nu}
	\left[ 2\delta\Gamma_{LMN}
	+ \delta Z_{L;MN} + \delta Z_{L;NM}
	+ (R^{(0)}_{L'MLN}+R^{(0)}_{L'NLM})\delta Z^{L'}\right],
	\label{eqn:another-delta-K}
\end{eqnarray}
%======================================%
where $\delta Z_M=g^{(0)}_{MN}\delta Z^N$, and 
$R^{(0)}_{L'MLN}$ is the Riemann tensor of the unperturbed metric
$g^{(0)}_{MN}$. To obtain this expression, we have used the following
identity for an arbitrary $D$-vector $V^M$. 
%============< EQUATION >==============%
%
\begin{equation}
 e^{(0)M}_{\mu}e^{(0)N}_{\nu}\left[
	{\cal L}_V{\cal L}_{n^{(0)}}g^{(0)}_{MN}
	- {\cal L}_{W[V]}g^{(0)}_{MN}
	+ n^{(0)L}(V_{L;MN}+V_{L;NM}
	+R_{L'MLN}V^{L'}+R_{L'NLM}V^{L'})\right]=0,
\end{equation}
%======================================%
where $V_M=g^{(0)}_{MN}V^N$ and 
%============< EQUATION >==============%
%
\begin{equation}
 W^M[V] = \epsilon n^{(0)M}n^{(0)N}n^{(0)L}V_{N;L}
	- e^{(0)M}_{\mu}q^{(0)\mu\nu}n^{(0)}_N
	{\cal L}_V e^{(0)N}_{\nu}. 
\end{equation}
%======================================%
The reason why we had to extend $\delta Z^M$ off
$\Sigma^{(0)}$ is only because we had to make covariant derivatives of 
$\delta Z^M$ well-defined on $\Sigma^{(0)}$ in this expression. The
value of $\delta K_{\mu\nu}$ is, off course, independent of the way of 
extension off $\Sigma^{(0)}$, since this expression is equivalent to 
the previous expression (\ref{eqn:delta-K-formal}) and no assumptions
with respect to properties of the extension have been needed. Off
course, the expression (\ref{eqn:another-delta-K}) holds for any
choices of the one-parameter family of hypersurfaces
(\ref{eqn:Sigma-varphi-def}), since no assumptions have been needed
with respect to the one-parameter family of hypersurfaces so far.

If we assume some properties with respect to the one-parameter family
of hypersurfaces (\ref{eqn:Sigma-varphi-def}), we can obtain other
expressions of $\delta K_{\mu\nu}$. One of them is given in
appendix~\ref{app:another-delta-K}, although we will not use it in the 
main body of this paper.

\subsection{Two kinds of gauge transformations}

%%%%% Invariance under spacetime coordinate transformation

Now let us investigate gauge transformation of $\delta q_{\mu\nu}$ and 
$\delta K_{\mu\nu}$. We have two kinds of gauge transformations. The
first one is the infinitesimal version of the $D$-coordinate
transformation (\ref{eqn:coord-tr}),
%============< EQUATION >==============%
%
\begin{equation}
 x^M \to {x'}^M = x^M+\bar{\xi}^M(x).
	\label{eqn:x->x'}
\end{equation}
%======================================%
We call it {\it $D$-gauge transformation}. Under the $D$-gauge 
transformation, $\delta g_{MN}$ and $\delta Z^M$ transform as 
%============< EQUATION >==============%
%
\begin{eqnarray}
 \delta g_{MN} & \to &
	\delta g_{MN} -\bar{\xi}_{M;N} -\bar{\xi}_{N;M},
	\nonumber\\
 \delta Z^M & \to & \delta Z^M + \bar{\xi}^M.
	\label{eqn:D-gauge-tr}
\end{eqnarray}
%======================================%
Hence, by using the expressions (\ref{eqn:delta-q}) and
(\ref{eqn:another-delta-K}), it is easy to show that 
$\delta q_{\mu\nu}$ and $\delta K_{\mu\nu}$ are invariant under the
$D$-gauge transformation. Off course, the $D$-gauge transformation can
be taken independently in the regions ${\cal M}_+$ and ${\cal M}_-$.

%%%%% Transformation under reparameterization

The second kind of gauge transformation is the infinitesimal version
of the $(D-1)$-coordinate transformation
(\ref{eqn:reparametrization}), 
%============< EQUATION >==============%
%
\begin{equation}
 y^{\mu} \to {y'}^{\mu} = y^{\mu}+\bar{\zeta}^{\mu}(y).
	\label{eqn:y->y'}
\end{equation}
%======================================%
We call it {\it $(D-1)$-gauge transformation}. Under the $(D-1)$-gauge 
transformation, 
%============< EQUATION >==============%
%
\begin{eqnarray}
 \left.\delta g_{MN}\right|_{x=Z^{0}(y)} 
	& \to & \left.\delta g_{MN}\right|_{x=Z^{0}(y)},
	\nonumber\\
 \delta Z^M & \to & \delta Z^M - \bar{\zeta}^{\mu}e^{(0)M}_{\mu}. 
\end{eqnarray}
%======================================%
Hence, by using the expressions (\ref{eqn:delta-q}) and
(\ref{eqn:delta-K-formal}), it is easy to show that 
$\delta q_{\mu\nu}$ and $\delta K_{\mu\nu}$ transform as follows under
the $(D-1)$-gauge transformation. 
%============< EQUATION >==============%
%
\begin{eqnarray}
 \delta q_{\mu\nu} & \to & \delta q_{\mu\nu} 
	- \bar{\cal L}_{\bar{\zeta}}q^{(0)}_{\mu\nu}, \nonumber\\
 \delta K_{\mu\nu} & \to & \delta K_{\mu\nu} 
	- \bar{\cal L}_{\bar{\zeta}}K^{(0)}_{\mu\nu},
\end{eqnarray}
%======================================%
where $\bar{\cal L}$ denotes the Lie derivative defined in the
$(D-1)$-dimensional manifold $\Sigma^{(0)}$, and we have used the
identity $[e^{(0)}_{\mu},e^{(0)}_{\nu}]^M=0$. Therefore, as expected,
$\delta q_{\mu\nu}$ and $\delta K_{\mu\nu}$ actually transform as
perturbations of $(D-1)$-tensors under the $(D-1)$-gauge
transformation.

\subsection{Junction condition}

%%%%% Junction condition

Finally, by decomposing the surface energy momentum $S_{\mu\nu}$ into
the unperturbed part $S^{(0)}_{\mu\nu}$ and the perturbation 
$\delta S_{\mu\nu}$, we obtain the following junction condition. 
%============< EQUATION >==============%
%
\begin{eqnarray}
 q^{(0)}_{\mu\nu+} & = & q^{(0)}_{\mu\nu-} \equiv q^{(0)}_{\mu\nu}, 
	\nonumber\\
 K^{(0)}_{\mu\nu +} - K^{(0)}_{\mu\nu -} & = & 
	-\kappa^2 \left(S^{(0)}_{\mu\nu}
	-\frac{1}{D-2}S^{(0)}q^{(0)}_{\mu\nu}\right)
\end{eqnarray}
%======================================%
for the background and 
%============< EQUATION >==============%
%
\begin{eqnarray}
 \delta q_{\mu\nu+} & = & \delta q_{\mu\nu-}, 
	\nonumber\\
 \delta\tilde{K}_{\mu\nu +} - \delta\tilde{K}_{\mu\nu -} & = &
	-\kappa^2\left(\delta\tilde{S}_{\mu\nu}
	-\frac{1}{D-2}\delta\tilde{S}q^{(0)}_{\mu\nu}\right)
	\label{eqn:perturbed-junction}
\end{eqnarray}
%======================================%
for the perturbation, where
%============< EQUATION >==============%
%
\begin{eqnarray}
 \delta\tilde{K}_{\mu\nu} & = & 
	\delta K_{\mu\nu} - \frac{1}{2}
	(K^{(0)\rho}_{\mu}\delta q_{\rho\nu}
	+K^{(0)\rho}_{\nu}\delta q_{\rho\mu}), \nonumber\\
 \delta\tilde{S}_{\mu\nu} & = & 
	\delta S_{\mu\nu} - \frac{1}{2}
	(S^{(0)\rho}_{\mu}\delta q_{\rho\nu}
	+S^{(0)\rho}_{\nu}\delta q_{\rho\mu}),
	\label{eqn:def-tildeK-tildeS}
\end{eqnarray}
%======================================%
and $S^{(0)}=q^{(0)\mu\nu}S^{(0)}_{\mu\nu}$ and 
$\delta\tilde{S}=q^{(0)\mu\nu}\delta\tilde{S}_{\mu\nu}$. 
Here, we have assumed that the unperturbed unit normal $n^{(0)}_M$ is
directed from ${\cal M}_-$ to ${\cal M}_+$. This perturbed junction
condition can be applied to general perturbations around an arbitrary
background. The quantity $\delta q_{\mu\nu}$ is given by
(\ref{eqn:delta-q}), while $\delta K_{\mu\nu}$ is given by
(\ref{eqn:delta-K-formal}) or (\ref{eqn:another-delta-K}). 
These quantities are invariant under the $D$-gauge
transformation, and they transform as perturbations of $(D-1)$-tensors
under the $(D-1)$-gauge transformation.

%======================================%
%<<<<<<<< Symmetric background >>>>>>>>%
%======================================%

\section{Symmetric background}
	\label{sec:symmetric-back}

In this section, as an application of the general scheme developed in
the previous section, we consider perturbation of the junction
condition around a background with the symmetry of a
$(D-2)$-dimensional constant curvature space. The perturbed junction
condition will be written in terms of doubly gauge-invariant variables
only.

%%%%% Background metric 

Let us consider a background metric with the symmetry of a
$(D-2)$-dimensional constant curvature space. 
%============< EQUATION >==============%
%
\begin{equation}
 g^{(0)}_{MN}dx^Mdx^N = 
	\gamma_{ab}dx^adx^b + r^2\Omega_{ij}dx^idx^j,
	\label{eqn:symmetric-g^0}
\end{equation}
%======================================%
where $\Omega_{ij}$ is the metric of the $(D-2)$-dimensional constant
curvature space with the curvature constant $K$, $\gamma_{ab}$ is a
$2$-dimensional metric. It is supposed that $\gamma_{ab}$ and $r$
depend only on the $2$-dimensional coordinates $\{x^a\}$. In this
background we have three kinds of covariant derivatives. The first one
is the $D$-dimensional covariant derivative compatible with
$g^{(0)}_{MN}$, which we represent by a semicolon. The second one is 
the $2$-dimensional covariant derivative $\nabla_a$ compatible with 
$\gamma_{ab}$. The final one is the $(D-2)$-dimensional covariant
derivative $D_i$ compatible with $\Omega_{ij}$. Relations among these
covariant derivatives are easily obtained. (For example, see
ref.~\cite{Mukohyama2}.) It is easy to show by explicit calculation
that Riemann tensor of the background metric $g^{(0)}_{MN}$ has the
following components. 
%============< EQUATION >==============%
%
\begin{eqnarray}
 R^{(0)ij}_{\qquad kl} & = & \left(
        \frac{K}{r^2}-\gamma^{ab}\partial_a\ln r \partial_b\ln r
        \right)(\delta^i_k \delta^j_l - \delta^i_l \delta^j_k),
        \nonumber\\
 R^{(0)i}_{\quad\ ajb} & = & -\delta^i_j
        (\nabla_a\nabla_b\ln r + \partial_a\ln r\partial_b\ln r), 
        \nonumber\\
 R^{(0)}_{abcd} & = & R^{(\gamma)}_{abcd}, 
\end{eqnarray}
%======================================%
and $R^{(0)}_{abci}=R^{(0)}_{abij}=R^{(0)}_{aijk}=0$, where
$R^{(\gamma)}_{abcd}$ is the Riemann tensor of $\gamma_{ab}$.

%%%%% \Sigma^{(0)}

Further, let us assume that the unperturbed hypersurface
$\Sigma^{(0)}$ also has the same symmetry. In this case, we can choose 
functions $Z^{(0)M}$ so that $Z^{(0)a}$ depends only on $y^0$ and that 
$Z^{(0)i}=y^i$. The unperturbed induced metric on $\Sigma^{(0)}$ is 
%============< EQUATION >==============%
%
\begin{equation}
 q^{(0)}_{\mu\nu}dx^{\mu}dx^{\nu} = 
	-\epsilon N^2dy^{0}dy^{0} + r^2\Omega_{ij}dy^idy^j, 
\end{equation}
%======================================%
where $N^2$ defined as follows and $r^2$ are written as functions of
$y^0$ by using $x^a=Z^{(0)a}(y^0)$, and $\Omega_{ij}$ is written in
terms of $\{y^{\mu}\}$ by using $x^i=y^i$. 
%============< EQUATION >==============%
%
\begin{eqnarray}
 N^2 & = & -\epsilon\gamma_{ab}e^ae^b, 
	\nonumber\\
 e^a & = & \partial Z^{(0)a}/\partial y^0.
\end{eqnarray}
%======================================%
As for the unit normal of $\Sigma^{(0)}$, $i$-components are zero and
$a$-components are determined by 
%============< EQUATION >==============%
%
\begin{eqnarray}
 e^a n^{(0)}_a  & = & 0,\nonumber\\
 n^{(0)a}n^{(0)}_a & = & \epsilon,
\end{eqnarray}
%======================================%
where $n^{(0)a}=\gamma^{ab}n^{(0)}_b$ and $\gamma^{ab}$ is the inverse 
of $\gamma_{ab}$. Correspondingly, the unperturbed extrinsic curvature
is 
%============< EQUATION >==============%
%
\begin{equation}
 K^{(0)}_{\mu\nu}dy^{\mu}dy^{\nu} = N^2{\cal K}dy^0dy^0 
	+ r^2\bar{\cal K}\Omega_{ij}dy^idy^j,
\end{equation}
%======================================%
where
%============< EQUATION >==============%
%
\begin{eqnarray}
 {\cal K} & = & N^{-2}e^a e^b \nabla_a n^{(0)}_b,\nonumber\\
 \bar{\cal K} & = & n^{(0)a}\partial_a\ln r.
\end{eqnarray}
%======================================%

%%%%% Unperturbed junction condition

Because of the symmetry assumed above, we can write the unperturbed
surface energy momentum tensor in the following form. 
%============< EQUATION >==============%
%
\begin{equation}
 S^{(0)}_{\mu\nu}dy^{\mu}dy^{\nu} = 
	N^2\rho dy^0dy^0 + r^2p\Omega_{ij}dy^idy^j,
\end{equation}
%======================================%
where $\rho$ and $p$ are functions of $y^0$ only. Thence, the
unperturbed junction condition is 
%============< EQUATION >==============%
%
\begin{eqnarray}
 N^2_+ - N^2_- & = & r^2_+ - r^2_- = 0, \nonumber\\
 {\cal K}_+ - {\cal K}_- & = & 
	-\kappa^2\left(\frac{n-1}{n}\rho+\epsilon p\right),
	\nonumber\\
 \bar{\cal K}_+ - \bar{\cal K}_- & = & 
	-\kappa^2\frac{\epsilon\rho}{n},
\end{eqnarray}
%======================================%
where $n=D-2$. 
Here, we have assumed that the unperturbed unit normal $n^{(0)}_a$ is
directed from ${\cal M}_-$ to ${\cal M}_+$.

\subsection{Perturbations and harmonic expansion}

%%%%% Perturbation

Perturbations $\delta q_{\mu\nu}$ and $\delta K_{\mu\nu}$ of the
induced metric and the extrinsic curvature around the background can be 
calculated by using the expressions (\ref{eqn:delta-q}) and
(\ref{eqn:another-delta-K}).
%============< EQUATION >==============%
%
\begin{eqnarray}
 \delta q_{00} & = & e^a e^b
	(\delta g_{ab}+2\nabla_a\delta Z_b),
	\nonumber\\
 \delta q_{0i} & = & e^a
	(\delta g_{ai}+\partial_i\delta Z_a+\partial_a\delta Z_i
	-2\delta Z_i\partial_a\ln r),
	\nonumber\\
 \delta q_{ij} & = & 
	\delta g_{ij}+D_i\delta Z_j+D_j\delta Z_i
	+2r^2\Omega_{ij}\delta Z^a\partial_a\ln r,
\end{eqnarray}
%======================================%
and 
%============< EQUATION >==============%
%
\begin{eqnarray}
 \delta K_{00} & = & \frac{\epsilon}{2}n^{(0)a}n^{(0)b}
	(\delta g_{ab}+\nabla_a\delta Z_b)N^2{\cal K}
	\nonumber\\
 & & 	- \frac{1}{2}n^{(0)a}e^be^c\left[
	(2\nabla_c\delta g_{ab}-\nabla_a\delta g_{bc})
	+2(\nabla_b\nabla_c\delta Z_a+R^{(\gamma)}_{dbac}\delta Z^d)
	\right],	\nonumber\\
 \delta K_{0i} & = & -\frac{1}{2}n^{(0)a}e^b(
	\partial_i\delta g_{ab}+\nabla_b\delta g_{ai}
	-\nabla_a\delta g_{bi}-2\delta g_{ai}\partial_b\ln r
	\nonumber\\
 & & 	+2\partial_i\nabla_b\delta Z_a 
	-2\partial_i\delta Z_a\partial_b\ln r
	-2\partial_b\delta Z_i\partial_a\ln r
	+4\delta Z_i\partial_a\ln r\partial_b\ln r),
	\nonumber\\
 \delta K_{ij} & = & \frac{\epsilon}{2}n^{(0)a}n^{(0)b}
	(\delta g_{ab}+\nabla_a\delta Z_b)r^2\bar{\cal K}\Omega_{ij}
	\nonumber\\
 & & 	- \frac{1}{2}n^{(0)a}\left[
	D_i\delta g_{aj}+D_j\delta g_{ai}
	+2r^2\Omega_{ij}\delta g_a^b\partial_b\ln r 
	-\partial_a\delta g_{ij}\right.
	\nonumber\\
 & & 	\left.+2D_iD_j\delta Z_a
	-2(D_i\delta Z_j+D_j\delta Z_i)\partial_a\ln r
	+2r^2\Omega_{ij}\nabla^b\delta Z_a\partial_b\ln r
	-\Omega_{ij}\delta Z^b\nabla_a\nabla_br^2\right],
\end{eqnarray}
%======================================%
where $\delta g_a^b=\gamma^{bc}\delta g_{ac}$. 
Here, we have used relations among three kinds of covariant
derivatives derived in ref.~\cite{Mukohyama2}. (The relations among 
covariant derivatives obtained in ref.~\cite{Mukohyama2} hold for
a general metric of the form (\ref{eqn:symmetric-g^0}) since the
relations are purely kinematical statements.)

%%%%% Harmonic expansion

Now, since the background has the symmetry of a $(D-2)$-dimensional
constant curvature space, it is convenient to expand perturbations by
using harmonics on the constant curvature space as follows. 
%============< EQUATION >==============%
%
\begin{eqnarray}
 \delta g_{MN}dx^Mdx^N & = & \sum_k\left[
	h_{ab}Ydx^adx^b 
	+ 2(h_{(T)a}V_{(T)i}+h_{(L)a}V_{(L)i})dx^adx^i
	\right.\nonumber\\
 & & 	\left.
	+ (h_{(T)}T_{(T)ij}+h_{(LT)}T_{(LT)ij}+
	h_{(LL)}T_{(LL)ij}+h_{(Y)}T_{(Y)ij})dx^idx^j\right],
	\nonumber\\
 \delta Z_Mdx^M & = & \sum_k\left[
	z_aYdx^a
	+ (z_{(T)}V_{(T)i}+z_{(L)}V_{(L)i})dx^i\right],
	\nonumber\\
 \delta S_{\mu\nu}dy^{\mu}dy^{\nu} & = & \sum_k\left[
	t_{00}Ydy^0dy^0
	+ 2(t_{(T)0}V_{(T)i}+t_{(L)0}V_{(L)i})dy^0dy^i
	\right.\nonumber\\
 & & 	\left.
	+ (t_{(T)}T_{(T)ij}+t_{(LT)}T_{(LT)ij}+
	t_{(LL)}T_{(LL)ij}+t_{(Y)}T_{(Y)ij})dy^idy^j\right],
	\label{eqn:harmonic-expansion}
\end{eqnarray}
%======================================%
where $Y$, $V_{(T,L)i}$ and $T_{(T,LT,LL,Y)ij}$ are scalar, vector and
tensor harmonics, respectively, and all coefficients are supposed to
depend only on the $2$-dimensional coordinates $\{x^a\}$. Hereafter,
$k$ denotes continuous ($K=0,-1$) or discrete ($K=1$) eigenvalues, and
we omit them in most cases. In this respect, the summation with
respect to $k$ should be understood as an integration for
$K=0,-1$. For definitions of these harmonics, see
Appendix~\ref{app:harmonics}. Here, we only mention that
$V_{(L)i}$, $T_{(LL)ij}$ and $T_{(Y)ij}$ are constructed from the
scalar harmonics $Y$ and that $T_{(LT)ij}$ is constructed from the
vector harmonics $V_{(T)i}$.

\subsection{$D$-gauge-invariant variables}

The next task might be to obtain the corresponding harmonic expansions
of $\delta q_{\mu\nu}$ and $\delta K_{\mu\nu}$. However, before doing
it, we shall introduce those linear combinations of perturbation
variables which are invariant under the $D$-gauge transformation
(\ref{eqn:x->x'}). Such linear combinations are usually 
called gauge-invariant variables. However, since we also have another
kind of gauge transformation (the $(D-1)$-gauge transformation
(\ref{eqn:y->y'})), we shall call those combinations 
{\it $D$-gauge-invariant variables}. It is expected that each 
coefficient of the harmonic expansions of $\delta q_{\mu\nu}$ and
$\delta K_{\mu\nu}$ can be written in terms of the $D$-gauge-invariant
variables only, since $\delta q_{\mu\nu}$ and $\delta K_{\mu\nu}$
should be invariant under the $D$-gauge transformation from the
general arguments in the previous section.

%%%%% $D$-gauge-invariant variables

The $D$-gauge transformation of $\delta g_{MN}$ and $\delta Z_M$ is
given by (\ref{eqn:D-gauge-tr}) and, by expanding the vector
$\bar{\xi}_M$ in terms of harmonics as 
%============< EQUATION >==============%
%
\begin{equation}
 \bar{\xi}_Mdx^M = \sum_k\left[\xi_aYdx^a + 
	(\xi_{(T)}V_{(T)i}+\xi_{(L)}V_{(L)i})dx^i\right],
\end{equation}
%======================================%
we get the following infinitesimal gauge transformation for the
expansion coefficients in Eqs.~(\ref{eqn:harmonic-expansion}). 
%============< EQUATION >==============%
%
\begin{eqnarray}
 h_{ab} & \to & 
	h_{ab} - \nabla_a\xi_b - \nabla_b\xi_a,
	\nonumber\\
 h_{(T)a} & \to & 
	h_{(T)a} - r^2\partial_a(r^{-2}\xi_{(T)}),
	\nonumber\\
 h_{(L)a} & \to & 
	h_{(L)a} - \xi_a - r^2\partial_a(r^{-2}\xi_{(L)}),
	\nonumber\\
 h_{(T)} & \to & 
	h_{(T)},
	\nonumber\\
 h_{(LT)} & \to & 
	h_{(LT)} - \xi_{(T)},
	\nonumber\\
 h_{(LL)} & \to & 
	h_{(LL)} - \xi_{(L)},
	\nonumber\\
 h_{(Y)} & \to & 
	h_{(Y)} - \gamma^{ab}\xi_a\partial_br^2 
	+ \frac{2k^2}{n}\xi_{(L)},
\end{eqnarray}
%======================================%
and
%============< EQUATION >==============%
%
\begin{eqnarray}
 z_a & \to & z_a + \xi_a, \nonumber\\
 z_{(T)} & \to & z_{(T)} + \xi_{(T)}, \nonumber\\
 z_{(L)} & \to & z_{(L)} + \xi_{(L)}.
\end{eqnarray}
%======================================%
Here, we have used relations among three kinds of covariant
derivatives derived in ref.~\cite{Mukohyama2}. 
Therefore, we can construct $D$-gauge-invariant variables
corresponding to perturbations of physical position of the
hypersurface $\Sigma$ 
%============< EQUATION >==============%
%
\begin{eqnarray}
 \phi_a & = & z_a + X_a, \nonumber\\
 \phi_{(L)} & = & z_{(L)} + h_{(LL)}, \nonumber\\
 \phi_{(T)} & = & z_{(T)} + h_{(LT)}
\end{eqnarray}
%======================================%
as well as $D$-gauge-invariant variables introduced in
ref.~\cite{Mukohyama2}, 
%============< EQUATION >==============%
%
\begin{eqnarray}
 F_{ab} & = & h_{ab}-\nabla_aX_b-\nabla_bX_a,
	\nonumber\\
 F & = & h_{(Y)} -X^a\partial_br^2+\frac{2k^2}{n}h_{(LL)},
	\nonumber\\
 F_{a} & = & h_{(T)a}-r^2\partial_a(r^{-2}h_{(LT)}),
	\nonumber\\
  F_{(T)} & = & h_{(T)},	\label{eqn:D-gauge-inv}
\end{eqnarray}
%======================================%
where $X_a=h_{(L)a}-r^2\partial_a(r^{-2}h_{(LL)})$. 
(In ref.~\cite{Mukohyama2} these $D$-gauge-invariant variables
(\ref{eqn:D-gauge-inv}) were simply called gauge-invariant variables
since only the $D$-gauge-transformation was considered in that paper.)

%%%%% \delta q, \delta K 

Now, it can actually be shown by using the harmonic expansion of 
$\delta g_{MN}$ and $\delta Z_M$ that $\delta q_{\mu\nu}$ and
$\delta K_{\mu\nu}$ are written in terms of $D$-gauge-invariant
variables only. 
%============< EQUATION >==============%
%
\begin{eqnarray}
 \delta q_{\mu\nu}dy^{\mu}dy^{\nu} & = & \sum_k\left[
	\sigma_{00}Ydy^0dy^0
	+ 2(\sigma_{(T)0}V_{(T)i}+\sigma_{(L)0}V_{(L)i})dy^0dy^i
	\right.\nonumber\\
 & & 	\left.
	+ (\sigma_{(T)}T_{(T)ij}+\sigma_{(LT)}T_{(LT)ij}+
	\sigma_{(LL)}T_{(LL)ij}+\sigma_{(Y)}T_{(Y)ij})dy^idy^j\right],
	\nonumber\\
 \delta K_{\mu\nu}dy^{\mu}dy^{\nu} & = & \sum_k\left[
	\kappa_{00}Ydy^0dy^0
	+ 2(\kappa_{(T)0}V_{(T)i}+\kappa_{(L)0}V_{(L)i})dy^0dy^i
	\right.\nonumber\\
 & & 	\left.
	+ (\kappa_{(T)}T_{(T)ij}+\kappa_{(LT)}T_{(LT)ij}+
	\kappa_{(LL)}T_{(LL)ij}+\kappa_{(Y)}T_{(Y)ij})dy^idy^j
	\right], 
\end{eqnarray}
%======================================%
where
%============< EQUATION >==============%
%
\begin{eqnarray}
 \sigma_{00} & = & e^a e^b (F_{ab}+2\nabla_a\phi_b),\nonumber\\
 \sigma_{(T)0} & = & e^a [F_{a}+r^2\partial_a(r^2\phi_{(T)})],
	\nonumber\\
 \sigma_{(L)0} & = & e^a [\phi_a+r^2\partial_a(r^2\phi_{(L)})],
	\nonumber\\
 \sigma_{(T)} & = & F_{(T)},
	\nonumber\\
 \sigma_{(LT)} & = & \phi_{(T)},
	\nonumber\\
 \sigma_{(LL)} & = & \phi_{(L)},
	\nonumber\\
 \sigma_{(Y)} & = & F+\phi^a\partial_ar^2-\frac{2k^2}{n}\phi_{(L)}, 
	\label{eqn:sigma-gauge-inv}
\end{eqnarray}
%======================================%
and 
%============< EQUATION >==============%
%
\begin{eqnarray}
 \kappa_{00} & = & \frac{\epsilon}{2}n^{(0)a}n^{(0)b}
	(F_{ab}+2\nabla_a\phi_b)N^2{\cal K}
	-\frac{1}{2}n^{(0)a}e^be^c(2\nabla_cF_{ab}-\nabla_{a}F_{bc}
	+2\nabla_b\nabla_c\phi_a+2R^{(\gamma)}_{dbac}\phi^d),
	\nonumber\\
 \kappa_{(T)0} & = & -\frac{1}{2}n^{(0)a}e^b\left[
	r^2\nabla_b(r^{-2}F_a)-\nabla_aF_b
	-\partial_ar^2\partial_b(r^{-2}\phi_{(T)})\right],
	\nonumber\\
 \kappa_{(L)0} & = & -\frac{1}{2}n^{(0)a}e^b\left[
	F_{ab}+2r\nabla_b(r^{-1}\phi_a)
	-\partial_ar^2\partial_b(r^{-2}\phi_{(L)})\right],
	\nonumber\\
 \kappa_{(T)} & = & \frac{1}{2}n^{(0)a}(\partial_aF_{(T)})
	\nonumber\\
 \kappa_{(LT)} & = & -\frac{1}{2}n^{(0)a}
	(F_a-2\phi_{(T)}\partial_a\ln r)
	\nonumber\\
 \kappa_{(LL)} & = & -\frac{1}{2}n^{(0)a}
	(\phi_a-2\phi_{(L)}\partial_a\ln r)
	\nonumber\\
 \kappa_{(Y)} & = & \frac{\epsilon}{4}n^{(0)a}n^{(0)b}n^{(0)c}
	(F_{ab}+2\nabla_a\phi_b)\partial_cr^2
	\nonumber\\
 & & 	-\frac{1}{2}n^{(0)a}\left(F_a^b\partial_br^2-\partial_aF
	+\nabla^b\phi_a\partial_br^2-\phi^b\nabla_a\nabla_br^2
	-\frac{2k^2}{n}\phi_a 
	+\frac{4k^2}{n}\phi_{(L)}\partial_a\ln r\right),
	\label{eqn:kappa's}
\end{eqnarray}
%======================================%
where $\phi^a=\gamma^{ab}\phi_b$ and $F_a^b=F_{ac}\gamma^{cb}$. 
Correspondingly, the quantities $\tilde{K}_{\mu\nu}$ and
$\tilde{S}_{\mu\nu}$, which are defined by
(\ref{eqn:def-tildeK-tildeS}) and which play important roles in the
perturbed junction condition (\ref{eqn:perturbed-junction}), are
expanded as 
%============< EQUATION >==============%
%
\begin{eqnarray}
 \delta \tilde{K}_{\mu\nu}dy^{\mu}dy^{\nu} & = & \sum_k\left[
	\tilde{\kappa}_{00}Ydy^0dy^0
	+ 2(\tilde{\kappa}_{(T)0}V_{(T)i}
	+\tilde{\kappa}_{(L)0}V_{(L)i})dy^0dy^i
	\right.\nonumber\\
 & & 	\left.
	+ (\tilde{\kappa}_{(T)}T_{(T)ij}
	+\tilde{\kappa}_{(LT)}T_{(LT)ij}
	+\tilde{\kappa}_{(LL)}T_{(LL)ij}
	+\tilde{\kappa}_{(Y)}T_{(Y)ij})dy^idy^j\right],
	\nonumber\\
 \delta \tilde{S}_{\mu\nu}dy^{\mu}dy^{\nu} & = & \sum_k\left[
	\tilde{t}_{00}Ydy^0dy^0
	+ 2(\tilde{t}_{(T)0}V_{(T)i}
	+\tilde{t}_{(L)0}V_{(L)i})dy^0dy^i
	\right.\nonumber\\
 & & 	\left.
	+ (\tilde{t}_{(T)}T_{(T)ij}
	+\tilde{t}_{(LT)}T_{(LT)ij}
	+\tilde{t}_{(LL)}T_{(LL)ij}
	+\tilde{t}_{(Y)}T_{(Y)ij})dy^idy^j\right],
\end{eqnarray}
%======================================%
where $\tilde{\kappa}$'s are defined by 
%============< EQUATION >==============%
%
\begin{eqnarray}
 \tilde{\kappa}_{00} & = & \kappa_{00}+\epsilon{\cal K}\sigma_{00},
	\nonumber\\
 \tilde{\kappa}_{(T)0} & = & \kappa_{(T)0}
	-\frac{1}{2}(\bar{\cal K}-\epsilon{\cal K})\sigma_{(T)0},
	\nonumber\\
 \tilde{\kappa}_{(L)0} & = & \kappa_{(L)0}
	-\frac{1}{2}(\bar{\cal K}-\epsilon{\cal K})\sigma_{(L)0},
	\nonumber\\
 \tilde{\kappa}_{(T)} & = & \kappa_{(T)}-\bar{\cal K}\sigma_{(T)},
	\nonumber\\
 \tilde{\kappa}_{(LT)} & = & 
	\kappa_{(LT)}-\bar{\cal K}\sigma_{(LT)},
	\nonumber\\
 \tilde{\kappa}_{(LL)} & = & 
	\kappa_{(LL)}-\bar{\cal K}\sigma_{(LL)},
	\nonumber\\
 \tilde{\kappa}_{(Y)} & = & 
	\kappa_{(Y)}-\bar{\cal K}\sigma_{(Y)},
	\nonumber\\
	\label{eqn:tilde-kappa's}
\end{eqnarray}
%======================================%
and $\tilde{t}$'s are defined by the same relations with
($\tilde{\kappa}$'s, $\kappa$'s, ${\cal K}$, $\bar{\cal K}$) replaced
by ($\tilde{t}$'s, $t$'s, $\rho$, $p$).

\subsection{$(D-1)$-gauge-invariant variables}

As already stressed several times, we have two kinds of independent
gauge transformations: $D$-gauge transformation and $(D-1)$-gauge
transformation. As for the $D$-gauge transformation, we have defined
$D$-gauge-invariant variables as quantities invariant under that and
have shown that $\delta q_{\mu\nu}$ and $\delta K_{\mu}$ (and thus
$\delta\tilde{K}_{\mu\nu}$) are written in terms of the
$D$-gauge-invariant variables only. Now let us investigate the second:
the $(D-1)$-gauge transformation (\ref{eqn:y->y'}). In what follows,
we shall construct those linear combinations of perturbations which
are invariant under that, and shall call them 
{\it $(D-1)$-gauge-invariant variables}. It is evident that a
$D$-gauge-invariant variable is not necessary a
$(D-1)$-gauge-invariant variable. This is because the $(D-1)$-gauge 
transformation is not a part of the $D$-gauge transformation. In
particular, the expansion coefficients $\sigma$'s, $\tilde{\kappa}$'s,
and $\tilde{t}$'s are not $(D-1)$-gauge-invariant variables although
they are $D$-gauge-invariant variables.

%%%%% f's

First, $\sigma$'s transform as follows under the $(D-1)$-gauge
transformation. 
%============< EQUATION >==============%
%
\begin{eqnarray}
 \sigma_{00} & \to & \sigma_{00} 
	- 2N\frac{\partial}{\partial y^0}(N^{-1}\zeta_0), 
	\nonumber\\
 \sigma_{(T)0} & \to & \sigma_{(T)0} 
	- r^2\frac{\partial}{\partial y^0}(r^{-2}\zeta_{(T)}), 
	\nonumber\\
 \sigma_{(L)0} & \to & \sigma_{(L)0} - \zeta_0
	- r^2\frac{\partial}{\partial y^0}(r^{-2}\zeta_{(L)}), 
	\nonumber\\
 \sigma_{(T)} & \to & \sigma_{(T)},
	\nonumber\\
 \sigma_{(LT)} & \to & \sigma_{(LT)} - \zeta_{(T)},
	\nonumber\\
 \sigma_{(LL)} & \to & \sigma_{(LL)} - \zeta_{(L)},
	\nonumber\\
 \sigma_{(Y)} & \to & \sigma_{(Y)} 
	+ \epsilon N^{-2}\zeta_0\frac{\partial r^2}{\partial y^0}
	+ \frac{2k^2}{n}\zeta_{(L)},
\end{eqnarray}
%======================================%
where we have expanded
$\bar{\zeta}_{\mu}=q^{(0)}_{\mu\nu}\bar{\zeta}^{\nu}$ as 
%============< EQUATION >==============%
%
\begin{equation}
 \bar{\zeta}_{\mu}dy^{\mu} = \sum_k \left[
	\zeta_0Ydy^0 + (\zeta_{(T)}V_{(T)i}+\zeta_{(L)}V_{(L)i})dy^i
	\right].
\end{equation}
%======================================%
Therefore, we can construct $(D-1)$-gauge-invariant variables
$f_{00}$, $f$, $f_{0}$ and $f_{(T)}$ by 
%============< EQUATION >==============%
%
\begin{eqnarray}
 f_{00} & = & \sigma_{00} 
	- 2N\frac{\partial}{\partial y^0}(N^{-1}\chi), 
	\nonumber\\
 f & = & \sigma_{(Y)} 
	+ \epsilon N^{-2}\chi\frac{\partial r^2}{\partial y^0}
	+ \frac{2k^2}{n}\sigma_{(LL)},
	\nonumber\\
 f_0 & = & \sigma_{(T)0} 
	- r^2\frac{\partial}{\partial y^0}(r^{-2}\sigma_{(LT)}),
	\nonumber\\
 f_{(T)} & = & \sigma_{(T)},
\end{eqnarray}
%======================================%
where $\chi=
\sigma_{(L)0}-r^2\partial(r^{-2}\sigma_{(LL)})/\partial y^{0}$. 
It is evident that these $(D-1)$-gauge-invariant variables can be
written in terms of $D$-gauge-invariant variables since $\sigma$'s 
have already been written by $D$-gauge-invariant variables in 
(\ref{eqn:sigma-gauge-inv}). In fact, we can show that 
%============< EQUATION >==============%
%
\begin{eqnarray}
 f_{00} & = & e^ae^bF_{ab} - 2N^2{\cal K}n^a\phi_a,
	\nonumber\\
 f & = & F + \epsilon n^an^b\phi_a\partial_b r^2
	\nonumber\\
 f_0 & = & e^aF_a
	\nonumber\\
 f_{(T)} & = & F_{(T)}. 
	\label{eqn:f's-doubly-inv}
\end{eqnarray}
%======================================%

%%%%% \tau's

Next, $\tilde{t}$'s transform as follows under the $(D-1)$-gauge
transformation. 
%============< EQUATION >==============%
%
\begin{eqnarray}
 \tilde{t}_{00} & \to & \tilde{t}_{00} 
	+ \epsilon \zeta_0\frac{\partial\rho}{\partial y^0},
	\nonumber\\
 \tilde{t}_{(T)0} & \to & \tilde{t}_{(T)0} 
	-\frac{1}{2}(p+\epsilon\rho)r^2
	\frac{\partial}{\partial y^0}(r^{-2}\zeta_{(T)}),
	\nonumber\\
 \tilde{t}_{(L)0} & \to & \tilde{t}_{(L)0} 
	+\frac{1}{2}(p+\epsilon\rho)\left[\zeta_0-r^2
	\frac{\partial}{\partial y^0}(r^{-2}\zeta_{(L)})\right],
	\nonumber\\
 \tilde{t}_{(T)} & \to & \tilde{t}_{(T)},
	\nonumber\\
 \tilde{t}_{(LT)} & \to & \tilde{t}_{(LT)},
	\nonumber\\
 \tilde{t}_{(LL)} & \to & \tilde{t}_{(LL)},
	\nonumber\\
 \tilde{t}_{(Y)} & \to & \tilde{t}_{(Y)}
	+\epsilon N^{-2}r^2\zeta_0\frac{\partial\rho}{\partial y^0}. 
\end{eqnarray}
%======================================%
Correspondingly, we can define the following set of
$(D-1)$-gauge-invariant variables. 
%============< EQUATION >==============%
%
\begin{eqnarray}
 \tau_{00} & = & \tilde{t}_{00}
	+ \epsilon\chi\frac{\partial\rho}{\partial y^0},
	\nonumber\\
 \tau_{(T)0} & = & \tilde{t}_{(T)0}
	- \frac{1}{2}(p+\epsilon\rho)\sigma_{(T)0},
	\nonumber\\
 \tau_{(L)0} & = & \tilde{t}_{(L)0}
	+ \frac{1}{2}(p+\epsilon\rho)
	\left[\chi 
	- r^2\frac{\partial}{\partial y^0}(r^{-2}\sigma_{LL})\right], 
	\nonumber\\
 \tau_{(T)} & = & \tilde{t}_{(T)}
	\nonumber\\
 \tau_{(LT)} & = & \tilde{t}_{(LT)}
	\nonumber\\
 \tau_{(LL)} & = & \tilde{t}_{(LL)}
	\nonumber\\
 \tau_{(Y)} & = & \tilde{t}_{(Y)}
	+ \epsilon N^{-2}r^2\chi\frac{\partial p}{\partial y^0}.
	\label{eqn:tau's-def}
\end{eqnarray}
%======================================%

%%%%% k's

Thirdly, $(D-1)$-gauge transformations of $\tilde{\kappa}$'s are the
same as those of $\tilde{t}$'s with ($\tilde{t}$'s, $\rho$, $p$)
replaced by ($\tilde{\kappa}$'s, ${\cal K}$, $\bar{\cal K}$). Thus, we
can construct $(D-1)$-gauge-invariant variables as follows. 
%============< EQUATION >==============%
%
\begin{eqnarray}
 k_{00} & = & \tilde{\kappa}_{00}
	+ \epsilon\chi\frac{\partial{\cal K}}{\partial y^0},
	\nonumber\\
 k_{(T)0} & = & \tilde{\kappa}_{(T)0}
	- \frac{1}{2}(\bar{\cal K}+\epsilon{\cal K})\sigma_{(T)0},
	\nonumber\\
 k_{(L)0} & = & \tilde{\kappa}_{(L)0}
	+ \frac{1}{2}(\bar{\cal K}+\epsilon{\cal K})
	\left[\chi 
	- r^2\frac{\partial}{\partial y^0}(r^{-2}\sigma_{LL})\right], 
	\nonumber\\
 k_{(T)} & = & \tilde{\kappa}_{(T)}
	\nonumber\\
 k_{(LT)} & = & \tilde{\kappa}_{(LT)}
	\nonumber\\
 k_{(LL)} & = & \tilde{\kappa}_{(LL)}
	\nonumber\\
 k_{(Y)} & = & \tilde{\kappa}_{(Y)}
	+ \epsilon N^{-2}r^2\chi\frac{\partial p}{\partial y^0}.
\end{eqnarray}
%======================================%
It is evident that these $(D-1)$-gauge-invariant variables can be
written in terms of $D$-gauge-invariant variables since 
$\tilde{\kappa}$'s and $\sigma$'s have already been written by
$D$-gauge-invariant variables in (\ref{eqn:sigma-gauge-inv}),
(\ref{eqn:kappa's}) and (\ref{eqn:tilde-kappa's}). Explicit
expressions are as follows. 
%============< EQUATION >==============%
%
\begin{eqnarray}
 k_{00} & = & -\frac{1}{2}n^{(0)a}e^be^c
	(2\nabla_cF_{ab}-\nabla_aF_{bc})
	+\frac{\epsilon}{2}(n^{(0)a}n^{(0)b}+2N^{-2}e^ae^b)
	F_{ab}N^2{\cal K}
	\nonumber\\
 & &	-N e^b\partial_b\left[ N^{-1}e^c\partial_c(n^{(0)a}\phi_a)
	\right]
	+\frac{1}{2}N^2(\epsilon R^{(\gamma)}+2{\cal K}^2)
	n^{(0)a}\phi_a, 
	\nonumber\\
 k_{(T)0} & = & -\frac{1}{2}r^2n^{(0)a}e^b\left[
	\nabla_b(r^{-2}F_a)-\nabla_a(r^{-2}F_b)\right],
	\nonumber\\
 k_{(L)0} & = & -\frac{1}{2}n^{(0)a}e^bF_{ab}
	-re^a\nabla_a(r^{-1}n^{(0)b}\phi_b),
	\nonumber\\
 k_{(T)} & = & \frac{1}{2}r^2n^{(0)a}\partial_a(r^{-2}F_{(T)}),
	\nonumber\\
 k_{(LT)} & = & -\frac{1}{2}n^{(0)a}F_a,
	\nonumber\\
 k_{(LL)} & = & -\frac{1}{2}n^{(0)a}\phi_a
	\nonumber\\
 k_{(Y)} & = & \frac{\epsilon}{4} F_{ab}n^{(0)a}
	(2N^{-2}e^be^c-n^{(0)b}n^{(0)c})\partial_cr^2
	+ \frac{1}{2}r^2n^{(0)a}\partial_a(r^{-2}F)
	\nonumber\\
 & &	+ \frac{\epsilon}{2}N^{-2}
	e^b\partial_b(n^{(0)a}\phi_a)e^c\partial_cr^2
	+ \left(\epsilon r^2n^{(0)b}n^{(0)c}\nabla_b\nabla_c\ln r 
	+\frac{k^2}{n}\right)n^{(0)a}\phi_a,
	\label{eqn:k's-doubly-inv}
\end{eqnarray}
%======================================%
where $R^{(\gamma)}$ is the Ricci scalar of $\gamma_{ab}$.

\subsection{Doubly gauge-invariant junction condition}

%%%%% Doubly gauge-invariant junction conditions

Now we can write down junction conditions for doubly gauge-invariant
variables $f$'s, $\tau$'s and $k$'s, where $f$'s are given by
(\ref{eqn:f's-doubly-inv}), $\tau$'s are given by
(\ref{eqn:tau's-def}), and $k$'s are given by
(\ref{eqn:k's-doubly-inv}). By {\it doubly gauge-invariant variables},
we mean variables invariant under both $D$-gauge transformation and
$(D-1)$-gauge transformation.

%%%%% Three categories of perturbations

In ref.~\cite{Mukohyama2}, a classificatory criterion was introduced
for gravitational perturbations around a maximally symmetric
spacetime. Namely, expansion coefficients of harmonics $Y$, $V_{(L)i}$,
$T_{(LL)ij}$ and $T_{(Y)ij}$ are called {\it scalar perturbations}; 
expansion coefficients of harmonics $V_{(T)i}$ and $T_{(LT)ij}$ are
called {\it vector perturbations}; 
expansion coefficients of harmonics $T_{(T)ij}$ are called 
{\it tensor perturbations}. This classification can be applied to the
situation in this paper, too. From the orthogonality between different
kinds of harmonics (see Appendix B of ref.~\cite{Mukohyama2}), it is
evident that perturbations belonging to different categories should be 
decoupled from each other. We can easily confirm this general
conclusion explicitly from eqs.~(\ref{eqn:f's-doubly-inv}), 
(\ref{eqn:tau's-def}), and (\ref{eqn:k's-doubly-inv}). 
Therefore, in what follows, we can write down the doubly
gauge-invariant junction conditions for each category independently.

%%%%% Scalar perturbations

First, the junction condition for scalar perturbations is
%============< EQUATION >==============%
%
\begin{eqnarray}
 f_{00+} - f_{00-} & = & f_+ - f_- = 0,	\nonumber\\
 k_{00+} - k_{00-} & = & -\kappa^2\left(
	\frac{n-1}{n}\tau_{00}+\epsilon N^2r^{-2}\tau_{(Y)}\right),
	\nonumber\\
 k_{(Y)+} - k_{(Y)-} & = & -\kappa^2
	\frac{\epsilon r^2}{nN^2}\tau_{00},
	\nonumber\\
 k_{(L)0+} - k_{(L)0-} & = & -\kappa^2\tau_{(L)0},
	\nonumber\\
 k_{(LL)+} - k_{(LL)-} & = & -\kappa^2\tau_{(LL)}.
	\label{eqn:junc-cond-scalar}
\end{eqnarray}
%======================================%

%%%%% Vector perturbations

Next, the junction condition for vector perturbations is 
%============< EQUATION >==============%
%
\begin{eqnarray}
 f_{0+} - f_{0-} & = & 0, \nonumber\\
 k_{(T)0+} - k_{(T)0-} & = & -\kappa^2\tau_{(T)0},
	\nonumber\\
 k_{(LT)+} - k_{(LT)-} & = & -\kappa^2\tau_{(LT)}. 
	\label{eqn:junc-cond-vector}
\end{eqnarray}
%======================================%

%%%%% Tensor perturbations

Finally, the junction condition for tensor perturbations is 
%============< EQUATION >==============%
%
\begin{eqnarray}
 f_{(T)+} - f_{(T)-} & = & 0,
	\nonumber\\
 k_{(T)+} - k_{(T)-} & = & -\kappa^2\tau_{(T)}. 
	\label{eqn:junc-cond-tensor}
\end{eqnarray}
%======================================%

%======================================%
%<<<<<<< Summary and discussion >>>>>>>%
%======================================%

\section{Summary and Discussion}
	\label{sec:summary}

%%%%% Summary

We have investigated perturbation of Israel's junction condition
across a timelike or spacelike hypersurface $\Sigma$ in a
$D$-dimensional spacetime. First, we formulated a general scheme of
perturbation of the junction condition around an arbitrary background
in a doubly covariant way. The junction condition is given by
(\ref{eqn:perturbed-junction}), where the quantity $\delta q_{\mu\nu}$ 
is given by (\ref{eqn:delta-q}) and $\delta K_{\mu\nu}$ is given by
(\ref{eqn:delta-K-formal}) or (\ref{eqn:another-delta-K}). 
It was shown that $\delta q_{\mu\nu}$ and $\delta K_{\mu\nu}$ are
invariant under $D$-gauge transformation and that they transform as
perturbations of $(D-1)$-tensors under $(D-1)$-gauge
transformation. The $D$-gauge transformation is an infinitesimal
coordinate transformation of the spacetime, and the $(D-1)$-gauge
transformation is an infinitesimal reparameterization of $\Sigma$. It
is important that the latter is not a part of the former but they are
completely independent.

Next, as an application of the general formalism, we analyzed
perturbation of the junction condition around a symmetric background
which has the symmetry of a $(D-2)$-dimensional constant curvature
space. By applying the general formalism to the symmetric background,
we have obtained the doubly gauge invariant junction condition for
perturbations. Junction conditions for scalar, vector, and tensor
perturbations are given by (\ref{eqn:junc-cond-scalar}),
(\ref{eqn:junc-cond-vector}) and (\ref{eqn:junc-cond-tensor}),
respectively, where doubly gauge-invariant variables $f$'s, $\tau$'s
and $k$'s are given by (\ref{eqn:f's-doubly-inv}),
(\ref{eqn:tau's-def}), and (\ref{eqn:k's-doubly-inv}), respectively.

%%%%% Maximally symmetric case

Now let us discuss about a special case in which the background metric
$g^{(0)}_{MN}$ is maximally symmetric or, equivalently, of constant
curvature. (Curvature constants, or cosmological constants, in two 
regions ${\cal M}_+$ and ${\cal M}_-$ may be different.) This off
course includes the anti de-Sitter spacetime which can be used as a
bulk geometry of the brane-world cosmology. It was recently shown
first in ref.~\cite{Mukohyama2} that for a maximally symmetric
background the $D$-gauge-invariant variables $F_{ab}$ and $F$ are
described by a common scalar-type master variable $\Phi_{(S)}$ and
that the $D$-gauge-invariant variable $F_a$ is also described by
another scalar-type master variable $\Phi_{(V)}$. 
%============< EQUATION >==============%
%
\begin{eqnarray}
 r^{D-4}F_{ab} & = & \nabla_a\nabla_b\Phi_{(S)} 
	-\frac{D-3}{D-2}\nabla^2\Phi_{(S)}\gamma_{ab}
	-\frac{2(D-4)\Lambda}{(D-1)(D-2)}\Phi_{(S)}\gamma_{ab},
	\nonumber\\
 r^{D-6}F & = & \frac{1}{D-2}\left[\nabla^2\Phi_{(S)}
	+\frac{4\Lambda}{(D-1)(D-2)}\Phi_{(S)}\right],
	\nonumber\\
 r^{D-4}F_a & = & \epsilon_a^{\ b}\partial_b\Phi_{(V)},
	\label{eqn:master-variables}
\end{eqnarray}
%======================================%
where $e_a=\gamma_{ac}e^c$, and $\epsilon_{ab}$ is the Levi-Civita
tensor defined by
$\epsilon_{01}=-\epsilon_{10}=\sqrt{|\det\gamma_{ab}|}$ and 
$\epsilon_{00}=\epsilon_{11}=0$. 
(The Levi-Civita tensor is reduced to 
$\epsilon_{ab}=\epsilon N^{-1}(n^{(0)a}ae^b-e^an^{(0)b})$ on the
hypersurface $\Sigma_{0}$.) 
By substituting the relations (\ref{eqn:master-variables}) into
eqs.~(\ref{eqn:k's-doubly-inv}), we can obtain expressions for $k$'s
in terms of the master variables. The corresponding junction
conditions, combined with the wave equations for the master variables
given in ref.~\cite{Mukohyama2}, can be considered as basic equations
for perturbations of the brane-world cosmology. 
Actually, in the brane world scenario, if we assume the $Z_2$ symmetry
then the curvature constants in the two regions ${\cal M}_+$ and
${\cal M}_-$ should be the same and the junction conditions
(\ref{eqn:junc-cond-scalar}), (\ref{eqn:junc-cond-vector}) and
(\ref{eqn:junc-cond-tensor}) are reduced to  
%============< EQUATION >==============%
%
\begin{eqnarray}
 2k_{00+} & = & -\kappa^2\left(
	\frac{n-1}{n}\tau_{00}+\epsilon N^2r^{-2}\tau_{(Y)}\right),
	\nonumber\\
 2k_{(Y)+} & = & -\kappa^2
	\frac{\epsilon r^2}{nN^2}\tau_{00},
	\nonumber\\
 2k_{(L)0+} & = & -\kappa^2\tau_{(L)0},
	\nonumber\\
 2k_{(LL)+} & = & -\kappa^2\tau_{(LL)},
	\nonumber\\
 2k_{(T)0+} & = & -\kappa^2\tau_{(T)0},
	\nonumber\\
 2k_{(LT)+} & = & -\kappa^2\tau_{(LT)},
	\nonumber\\
 2k_{(T)+} & = & -\kappa^2\tau_{(T)}. 
\end{eqnarray}
%======================================%
These junction condition are equivalent to those obtained in
ref.~\cite{KIS}. (In appendix~\ref{app:relations}, some relations
between variables defined in the present paper and those in other
papers are given.) In future publications we would like to analyze
perturbations of the brane-world cosmology in detail by using the
basic equations.

%%%%% Black hole perturbation

If the brane-world scenario is realistic then it might also be
interesting to analyze black hole perturbations in the
brane-world. We may be able to expect that the brane-world scenario
might give non-standard predictions for gravitational waves emitted
from, say, a binary black hole system. For this purpose, off course,
we have to analyze background black hole solutions in the brane-world
scenario thoroughly. After that, the general doubly covariant
formalism developed in Sec.~\ref{sec:general} may be useful for the
analysis of perturbations.

%%%%%%%%%%%%%%%%%%%%%%%%%%%%%%%%%%%%%%%%%%%%%%%%%%%%%%%%%%%%%%%%%%%%
%%%%%%%%%%%%%%%%%%%%%%%%%%%%%%%%%%%%%%%%%%%%%%%%%%%%%%%%%%%%%%%%%%%%
% Acknowledgments
%%%%%%%%%%%%%%%%%%%%%%%%%%%%%%%%%%%%%%%%%%%%%%%%%%%%%%%%%%%%%%%%%%%%
%%%%%%%%%%%%%%%%%%%%%%%%%%%%%%%%%%%%%%%%%%%%%%%%%%%%%%%%%%%%%%%%%%%%
\begin{acknowledgments}

The author would like to thank Professor W. Israel for continuing
encouragement and discussions. He would be grateful to
Dr. T. Shiromizu, Professor M. Sasaki and Professor H. Kodama for
comments. This work was supported by the CITA National Fellowship and
the NSERC operating research grant. 

\end{acknowledgments}

%%%%%%%%%%%%%%%%%%%%%%%%%%%%%%%%%%%%%%%%%%%%%%%%%%%%%%%%%%%%%%%%%%%%
%%%%%%%%%%%%%%%%%%%%%%%%%%%%%%%%%%%%%%%%%%%%%%%%%%%%%%%%%%%%%%%%%%%%
% Appendix
%%%%%%%%%%%%%%%%%%%%%%%%%%%%%%%%%%%%%%%%%%%%%%%%%%%%%%%%%%%%%%%%%%%%
%%%%%%%%%%%%%%%%%%%%%%%%%%%%%%%%%%%%%%%%%%%%%%%%%%%%%%%%%%%%%%%%%%%%

\appendix

%======================================%
%<<<<<<< Another expression >>>>>>>>>>>%
%======================================%

\section{Another expression of $\delta K_{\mu\nu}$}
	\label{app:another-delta-K}

In the main body of this paper, no assumptions have been needed with 
respect to the one-parameter family of hypersurfaces
(\ref{eqn:Sigma-varphi-def}). In particular, the expression
(\ref{eqn:another-delta-K}) holds for any choices of the one-parameter
family of hypersurfaces. On the other hand, in order to obtain another
expression of $\delta K_{\mu\nu}$ which is similar to the expression
used in ref.~\cite{KIS}, we have to assume that
$\partial_{[M}n^{(0)}_{n]}=0$. Once this is assumed, we can obtain the
following expression for $\delta K_{\mu\nu}$. 
%============< EQUATION >==============%
%
\begin{equation}
 \delta K_{\mu\nu} = 
        \frac{\epsilon}{2}\delta g_{\perp\perp}K^{(0)}_{\mu\nu}
        - \delta\Gamma_{\perp\mu\nu}
        - \bar{\nabla}_{\mu}\bar{\nabla}_{\nu}\delta Z_{\perp}
        + \epsilon\delta Z_{\perp}
        (K^{(0)\rho}_{\mu}K^{(0)}_{\rho\nu}
        -R^{(0)}_{\perp\mu\perp\nu\perp})
        + \bar{\cal L}_{\delta Z_{\parallel}}K^{(0)}_{\mu\nu},
        \label{eqn:onemore-delta-K}
\end{equation}
%======================================%
where
%============< EQUATION >==============%
%
\begin{eqnarray}
 \delta g_{\perp\perp} & = & n^{(0)M}n^{(0)N}\delta g_{MN}, 
	\nonumber\\
 \delta\Gamma_{\perp\mu\nu} & = & 
	n^{(0)L}e^{(0)M}_{\mu}e^{(0)N}_{\nu}\delta\Gamma_{LMN},
	\nonumber\\
 R^{(0)}_{\perp\mu\perp\nu\perp} & = & 
	n^{(0)L'}e^{(0)M}_{\mu}n^{(0)L}e^{(0)N}_{\nu}
	R^{(0)}_{L'MLN}, \nonumber\\
 K^{(0)\rho}_{\mu} & = & K^{(0)}_{\mu\sigma}q^{(0)\sigma\rho},
	\nonumber\\
 \delta Z_{\perp} & = & 
	n^{(0)}_M\delta Z^M, \nonumber\\
 \delta Z^{\mu}_{\parallel} & = & 
	q^{(0)\mu\nu}e^{(0)M}_{\nu}g^{(0)}_{MN}\delta Z^N,
	\label{eqn:perp-parallel}
\end{eqnarray}
%======================================%
and $\bar{\nabla}$ denotes the $(D-1)$-dimensional covariant
derivative compatible with the unperturbed induced metric
$q^{(0)}_{\mu\nu}$ on $\Sigma^{(0)}$, $\bar{\cal L}$ denotes the Lie
derivative defined in the $(D-1)$-dimensional manifold $\Sigma^{(0)}$, and 
$R^{(0)}_{L'MLN}$ is the Riemann tensor of the unperturbed metric
$g^{(0)}_{MN}$. Note that all defined in
eqs.~(\ref{eqn:perp-parallel}) are $D$-scalars while they transform as 
$(D-1)$-scalars, $(D-1)$-vectors, or $(D-1)$-tensors under the
$(D-1)$-coordinate transformation. In particular, 
$\delta Z_{\parallel}^{\mu}$ is a $(D-1)$-vector-valued $D$-scalar.

In the expression (\ref{eqn:another-delta-K}) used in the main body of 
this paper, all derivatives are $D$-dimensional covariant derivatives
compatible with $g^{(0)}_{MN}$. Hence, it is easier to calculate
$\delta K_{\mu\nu}$ by using the expression
(\ref{eqn:another-delta-K}) than the expression 
(\ref{eqn:onemore-delta-K}).

%======================================%
%<<<<<<<<<<<<< Harmonics >>>>>>>>>>>>>>%
%======================================%

\section{Harmonics on constant-curvature space}
	\label{app:harmonics}

In this Appendix we give definitions of scalar, vector and tensor
harmonics on a $n$-dimensional constant-curvature space, where $n=D-2$ 
and $D$ is the spacetime dimension. Throughout this Appendix we will
use the notation that $\Omega_{ij}$ is the metric of the
constant-curvature space and that $D_i$ is the covariant derivative
compatible with $\Omega_{ij}$. For properties of these harmonics, see
Appendix B of ref.~\cite{Mukohyama2}.

\subsection{scalar harmonics}

The scalar harmonics is supposed to satisfy the following relations.
%============< EQUATION >==============%
%
\begin{eqnarray}
 D^2 Y + k^2Y & = & 0,	\nonumber\\
 \int d^nx\sqrt{\Omega}Y Y & = & \delta. 
\end{eqnarray}
%======================================%
Hereafter, $k^2$ denotes continuous eigenvalues for $K=0,-1$~\cite{VS} 
or discrete eigenvalues $k_l^2=l(l+n-1)$ ($l=0,1,\cdots$) for
$K=1$~\cite{RR}, and we omit them in most cases. In this respect, the 
delta $\delta$ in equations above and below represents Dirac's delta
function $\delta^n(k-k')$ for continuous eigenvalues and Kronecker's
delta $\delta_{ll'}\delta_{mm'}$ for discrete eigenvalues, where $m$
(and $m'$) denotes a set of integers. Correspondingly, in the
following arguments, a summation with respect to $k$ should be
understood as integration for $K=0,-1$.

\subsection{vector harmonics}

First, in general, a vector field $v_i$ can be decomposed as 
%============< EQUATION >==============%
%
\begin{equation} 
 v_i=v_{(T)i}+\partial_i f , 
\end{equation}
%======================================%
where $f$ is a function and $v_{(T)}$ is a transverse vector field:
%============< EQUATION >==============%
%
\begin{equation}
 D^iv_{(T)i}=0 .
\end{equation}
%======================================%

Thus, the vector field $v_i$ can be expanded by using the scalar
harmonics $Y$ and transverse vector harmonics $V_{(T)i}$ as 
%============< EQUATION >==============%
%
\begin{equation}
 V_i = \sum_k\left[c_{(T)}V_{(T)i}+c_{(L)}\partial_i Y\right],
	\label{eqn:dY+V}
\end{equation}
%======================================%
where $c_{(T)}$ and $c_{(L)}$ are constants depending on $k$, and the
transverse vector harmonics $V_{(T)i}$ is supposed to satisfy the
following relations.  
%============< EQUATION >==============%
%
\begin{eqnarray}
 D^2V_{(T)i} + k^2 V_{(T)i} & = & 0,	\nonumber\\
 D^iV_{(T)i} & = & 0,\nonumber\\
 \int d^nx\sqrt{\Omega}\Omega^{ij}V_{(T)i}V_{(T)j}
	& = & \delta,
\end{eqnarray}
%======================================%
where $k^2$ denotes continuous eigenvalues for $K=0,-1$ or discrete
eigenvalues $k_l^2=l(l+n-1)-1$ ($l=1,2,\cdots$) for $K=1$~\cite{RR},
and we omit them in most cases. From Eq.~(\ref{eqn:dY+V}), it is
convenient to define longitudinal vector harmonics $V_{(L)i}$ by  
%============< EQUATION >==============%
%
\begin{equation}
 V_{(L)i} \equiv \partial_i Y. 
\end{equation}
%======================================%

\subsection{Tensor harmonics}

First, in general, a symmetric second-rank tensor field $t_{ij}$ can
be decomposed as
%============< EQUATION >==============%
%
\begin{equation}
 t_{ij}=t_{(T)ij} + D_iv_j+D_jv_i + f\Omega_{ij},
\end{equation}
%======================================%
where $f$ is a function, $v_i$ is a vector field and $t_{(T)ij}$ is
a transverse traceless symmetric tensor field:
%============< EQUATION >==============%
%
\begin{eqnarray}
 t_{(T)i}^i & = & 0,\nonumber\\
 D^i t_{(T)ij} & = &0. 
	\label{eqn:trasverse-traceless}
\end{eqnarray}
%======================================%

Thus, the tensor field $t_{ij}$ can be expanded by using the vector
harmonics $V_{(T)}$ and $V_{(L)}$, and transverse traceless tensor 
harmonics $T_{(T)}$ as 
%============< EQUATION >==============%
%
\begin{equation}
 t_{ij} = \sum_{k}\left[
	c_{(T)}T_{(T)ij}+c_{(LT)}(D_iV_{(T)j}+D_jV_{(T)i})
	+ c_{(LL)}(D_iV_{(L)j}+D_jV_{(L)i})
	+ c_{(Y)}Y\Omega_{ij}\right],
	\label{eqn:dV+T}
\end{equation}
%======================================%
where $c_{(T)}$, $c_{(LT)}$, $c_{(LL)}$ and $c_{(Y)}$ are constants
depending on $k$, and the transverse tensor harmonics $T_{(T)}$ is
supposed to satisfy the following relations. 
%============< EQUATION >==============%
%
\begin{eqnarray}
 D^2 T_{(T)ij}+k^2T_{(T)ij} & = & 0,\nonumber\\ 
 T_{(T)i}^i & = & 0,\nonumber\\
 D^iT_{(T)ij} & = & 0,\nonumber\\
 \int d^nx\sqrt{\Omega}\Omega^{ik}\Omega^{jl}T_{(T)ij}T_{(T)kl} 
	& = & \delta, 
\end{eqnarray}
%======================================%
where $k^2$ denotes continuous eigenvalues for $K=0,-1$ or discrete
eigenvalues $k_l^2=l(l+n-1)-2$ ($l=2,3,\cdots$) for $K=1$~\cite{RR},
and we omit them in most cases. From Eq.~(\ref{eqn:dV+T}), it is
convenient to define tensor harmonics $T_{(LT)}$, $T_{(LL)}$, and
$T_{(Y)}$ by 
%============< EQUATION >==============%
%
\begin{eqnarray}
 T_{(LT)ij} & \equiv & D_iV_{(T)j}+D_jV_{(T)i}, \nonumber\\
 T_{(LL)ij} & \equiv & D_iV_{(L)j}+D_jV_{(L)i}
	-\frac{2}{n}\Omega_{ij}D^kV_{(L)k}	\nonumber\\
 & = & 2D_iD_jY+\frac{2}{n}k^2\Omega_{ij}Y,	\nonumber\\
 T_{(Y)ij} & \equiv & \Omega_{ij}Y. 
\end{eqnarray}
%======================================%

%======================================%
%<<<<<<<<<<<<< Relations >>>>>>>>>>>>>>%
%======================================%

\section{Relations to variables defined in other references}
	\label{app:relations}

For the sake of readers who like to compare the results of this paper
with those in other references, in this appendix, we shall give some
relations between variables defined in this paper and those in other
papers.

%%%%% Relations to Kodama-Sasaki variables

The $(D-1)$-gauge-invariant variables defined in this paper can be
related to gauge-invariant variables used in cosmology. Here, we give
relations between the $(D-1)$-gauge-invariant variables and those
gauge-invariant variables defined by Kodama and
Sasaki~\cite{Kodama-Sasaki}. 
(In ref.~\cite{Kodama-Sasaki}, $\epsilon=+$ and $N=r$. ) 
In the following, the subscript $(KS)$ represents a quantity defined
in ref.~\cite{Kodama-Sasaki}, and we use the relations 
$(Y$, $V_{(T)i}$, $V_{(L)i}$, $T_{(T)ij}$, $T_{(LT)ij}$, $T_{(LL)ij}$,
$T_{(Y)ij})=(Y$, $Y^{(1)}_{i}$, $-\sqrt{k^2}Y_{i}$, $Y^{(2)}_{ij}$,
$-2\sqrt{k^2}Y^{(1)}_{ij}$, $2k^2Y_{ij}$, $Y\Omega_{ij})_{(KS)}$. 
For scalar perturbations,
%============< EQUATION >==============%
%
\begin{eqnarray}
 N^{-2}f_{00} & = & -2\Psi_{(KS)}, \nonumber\\
 f & = & 2r^2\Phi_{(KS)}, \nonumber\\
 N^{-2}\tau_{00} & = & 
	\rho\Delta_{(KS)} 
	- N^{-1}\frac{\partial\rho}{\partial y^0}
	\frac{rV_{(KS)}}{-\sqrt{k^2}}, \nonumber\\
 \tau_{(Y)} & = & r^2(p\Gamma_{(KS)}+c^2_sN^{-2}\tau_{00}),
	\nonumber\\
 N^{-1}\tau_{(L)0} & = & -(p+\rho)
	\frac{rV_{(KS)}}{-\sqrt{k^2}}, \nonumber\\
 \tau_{(LL)} & = & r^2p \frac{\Pi_{(KS)}}{2k^2}, 
 \end{eqnarray}
%======================================%
where $c_s$ is the sound velocity defined by 
$c^2_s=(\partial p/\partial y^0)/(\partial\rho/\partial y^0)$.
Relations for vector perturbations are
%============< EQUATION >==============%
%
\begin{eqnarray}
 N^{-1}f_0 & = & r\sigma_{g(KS)}^{(1)}, \nonumber\\
 N^{-1}\tau_{(T)0} & = & (p+\rho)rV^{(1)}_{(KS)},
	\nonumber\\
 \tau_{(LT)} & = & r^2 p \frac{\Pi^{(1)}_{(KS)}}{-2\sqrt{k^2}}.
\end{eqnarray}
%======================================%
For tensor perturbations, 
%============< EQUATION >==============%
%
\begin{eqnarray}
 f_{(T)} & = & 2r^2H^{(2)}_{T(KS)}, \nonumber\\
 \tau_{(T)} & = & r^2 p \Pi^{(2)}_{(KS)}. 
\end{eqnarray}
%======================================%

%%%%% Relations to Kodama-Ishibashi-Seto variables

Next, let us give relations between the $D$-gauge-invariant variables
defined in this paper and those defined in ref.~\cite{KIS}. 
(In ref.~\cite{KIS}, $\epsilon=+$ and $N=1$. ) 
In the following, the subscript $(KIS)$ represents a quantity defined
in ref.~\cite{KIS}. 

%============< EQUATION >==============%
%
\begin{eqnarray}
 F_{ab} & = & F_{ab(KIS)},	\nonumber\\
 F & = & 2r^2 F_{(KIS)},	\nonumber\\
 F_a & = & rF_{a(KIS)},	\nonumber\\
 F_{(T)} & = & 2r^2H_{T(KIS)}
\end{eqnarray}
%======================================%

%%%%% Relations between Kodama-Sasaki and Kodama-Ishibashi-Seto 

Finally, we point out some relations between definitions in
ref.~\cite{Kodama-Sasaki} and ref.~\cite{KIS}. 
%============< EQUATION >==============%
%
\begin{eqnarray}
 p\Gamma_{(KS)} & = & \Gamma_{(KIS)},	\nonumber\\
 p\Pi_{(KS)} & = & \pi_{T(KIS)},	\nonumber\\
 p\Pi^{(1)}_{(KS)} & = & \pi_{T(KIS)},	\nonumber\\
 p\Pi^{(2)}_{(KS)} & = & \pi_{T(KIS)}.
\end{eqnarray}
%======================================%

%======================================%
%<<<<<<<<<<<< REFERENCES >>>>>>>>>>>>>>%
%======================================%

\end{document}